\documentclass[prc,aps,twocolumn,showpacs,preprintnumbers,eqsecnum,amsmath,amssymb,twocolomn]{revtex4-1}
\usepackage{graphicx}% Include figure files
\usepackage{epsfig}% Include figure files
\usepackage{dcolumn}% Align table columns on decimal point
\usepackage{bm}% bold math

\begin{document}
\title{Pygmy Quadrupole Resonance in Skin Nuclei}

\author{
 N.~Tsoneva$^{1,2}$, H.~Lenske$^{1}$}
\affiliation{
  $^1$Institut f\"ur Theoretische Physik, Universit\"at Gie\ss en,
  Heinrich-Buff-Ring 16, D-35392 Gie\ss en, Germany \\
$^2${Institute for Nuclear Research and Nuclear Energy, 1784 Sofia, Bulgaria}}

\begin{abstract}
 The electric quadrupole response is investigated theoretically by HFB and QPM calculations along the Sn isotopic chain with special emphasis on excitations above the first collective state and below the particle threshold. Depending on the asymmetry, additional strength clustering as a group of states similar to the known PDR mode is found. The spectral distributions and electric response functions are discussed. The transition densities of these Pygmy Quadrupole Resonances (PQR) states are closely related to the neutron excess and showing special features being compatible with an oscillation of the neutron skin against the nuclear core. These features may indicate a new nuclear quadrupole mode connected to the skin configuration.
\end{abstract}

\maketitle

%%%%%%%%%%%%%%%%%%%%%%%%%%%%%%%%%%%%%%%%%%%%
%% MAINMATTER
%%%%%%%%%%%%%%%%%%%%%%%%%%%%%%%%%%%%%%%%%%%%

\section{Introduction}

The new radioactive beam facilities allow to probe hitherto unaccessible regions of nuclear masses far from the valley of stability \cite{PDRrev:06}. The prevalence of one type of nucleons in medium and heavy nuclei may result in the formation of a nuclear skin, containing either pure proton or neutron matter, respectively. After the detection of halos in light nuclei \cite{b8,c19}, the skin phenomena adds another new aspect to the dynamics of strongly asymmetric nuclear matter, observable in medium and heavy exotic nuclei. Besides the genuine importance for nuclear structure physics, skin nuclei are of particular relevance for nuclear astrophysics \cite{Iocco,astro}. For example, the existence of low-energy excitation modes related to the neutron skins could influence the rapid (r) neutron capture process in low-energy (n,$\gamma$) reactions and, therefore, will affect directly the synthesis of heavy nuclei in stellar environments.

One of the most interesting results was the discovery of a new dipole mode at energies below and close to the particle emission threshold \cite{PDRrev:06}, observed in stable and unstable nuclei with isospin asymmetry $N/Z>1$. Typically that strength is found as a bunching of discrete $1^-$ states with very similar spectroscopic features. Although located below the particle threshold, the clustering of states in a narrow energy region resembles a resonance phenomena. This extra dipole strength was unexpected and could not be explained as just belonging to the low-energy part of the giant dipole resonance (GDR) \cite{Ts-Schw08}. Because of the unique character of the excitation mechanism the mode was named pygmy dipole resonance (PDR). The PDR is of high current interest and its properties are investigated by a large number of theoretical and experimental groups \cite{Govaert98,Nadia04,Paar,Kamer04,Sarchi04,Rye:02,Paar05,Volz06,Litv07,Bar08,Ts-Schw08,Ozel,Nadia08,Adrich:2005}.
Most of the approaches agree on the hypothesis that the PDR is an isospin-mixed mode due to excitations of identical particles, neutrons (protons) from the nuclear surface, corresponding to an oscillatory motion of the nuclear skin against the core \cite{Nadia08,Paar,Paar05,Kamer04,Sarchi04,Litv07,Bar08}. Despite remaining differences among the theoretical approaches (in particular between relativistic and non-relativistic calculations), there is agreement on the gross features of these excitations as a new nuclear mode, being closely related to and correlated with the size of the nuclear skin, as discussed in detail e.g. in refs. \cite{Adrich:2005,Nadia08,Cosel09}.

An obvious question, coming up immediately in this connection, is to what extent the presence of a neutron (proton) skin will affect also excitations of other multipolarities. Promising candidates are low-energy 2$^{+}$ states, in excess of the spectral distributions known from stable nuclei close to the valley of stability. Previous theoretical studies are found in \cite{VIsac92,Warner97,Yoko06}. Similar results were presented in our recent quasiparticle random phase approximation (QRPA) calculations in $^{120}$Sn \cite{NT09} where low-energy electric quadrupole strength, located well below the isoscalar giant quadrupole resonance (ISGQR) \cite{Hara} was observed. In order to clarify whether this is a general feature of the skin nuclei we have studies the quadrupole response along the Sn-isotopic chain.
This report contains results based on a Hartree-Fock-Bogoljubov (HFB) description of the ground state \cite{Hof98}
by using a phenomenological energy-density functional (EDF) \cite{Nadia04,Nadia08}. The excited states are calculated with QRPA theory, using the multiphonon description of the Quasiparticle-Phonon model (QPM) \cite{Sol76}.

\section{Theoretical model}

The building blocks of the model are given by the quasiparticle states obtained from a HFB description of the nuclear ground states and  coherent superpositions of two-quasiparticle states treated by QRPA theory. A detailed description of the approach is found in \cite{Nadia04,Nadia08}, here we give only a brief overview. 
The approach is based on the Hamiltonian
\begin{equation}
H=H_{MF}+H_{res} \quad ,
\label{hh}
\end{equation}
The mean-field part $H_{MF}$ defines the single particle properties and, as such, accounts for the ground state dynamics, including potentials and pairing interactions for protons and neutrons, respectively. That part is related to an EDF \cite{Nadia04,Nadia08} constructed such that also dynamical effects beyond mean-field can be taken into account. That goal is achieved in practice by using fully microscopic HFB potentials and pairing fields as input but performing a second step variation with scaled auxiliary potentials and pairing fields readjusted in a self-consistent manner such that nuclear binding energies and other ground state properties of relevance are closely reproduced.

The residual interactions contained in $H_{res}$ are treated separately. After having derived the ground state single quasiparticle spectra and wave functions the nuclear excited states are described by QRPA phonons of various multipolarities. Following the notation of \cite{Sol76}, the state operators are defined as a linear combination of two-quasiparticle creation and annihilation operators corresponding to excitations propagating forward and backward in time, respectively:
\begin{equation}
Q^{+}_{\lambda \mu i}=\frac{1}{2}{
\sum_{jj'}{ \left(\psi_{jj'}^{\lambda i}A^+_{\lambda\mu}(jj')
-\varphi_{jj'}^{\lambda i}\widetilde{A}_{\lambda\mu}(jj')
\right)}}.
\label{eq:StateOp}
\end{equation}
Here, $j\equiv{(nljm\tau)}$ is a single-particle proton or neutron state;
${A}^+_{\lambda \mu}$ and $\widetilde{A}_{\lambda \mu}$ are
proton and neutron two-quasiparticle creation and annihilation operators, respectively, with configuration amplitudes given by the time-forward and time-backward amplitudes
$\psi_{j_1j_2}^{\lambda i}$ and $\varphi_{j_1j_2}^{\lambda i}$ \cite{Sol76} ,
respectively. The operators are coupled to total
angular momentum $\lambda$ with projection $\mu$ by means of the
Clebsch-Gordan coefficients $C^{\lambda\mu}_{jmj'm'}=\left\langle
jmj'm'|\lambda\mu\right\rangle$. The phonons obey the QRPA equation of motion
\begin{equation}
\left[H,Q^+_\alpha\right]=E_\alpha Q^+_\alpha \quad ,
\label{ep}
\end{equation}
accounting now also for the residual interactions in the particle-hole and the particle-particle channels, respectively. By solving the eigenvalue problem, eq. \ref{ep}, we obtain the phonon excitation energies $E_\alpha$ and the state vectors, eq. \ref{eq:StateOp}.

\begin{figure}
\includegraphics[height=.18\textheight,width=.45\textwidth]{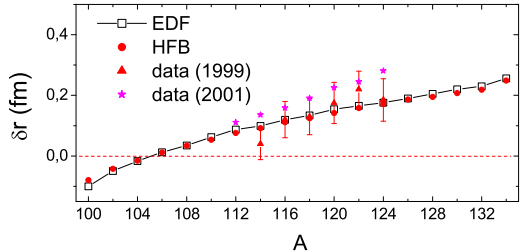}
\caption{EDF and HFB calculations of nuclear skin thicknesses $\delta r$   in comparison with skin thicknesses derived from charge exchange reactions by Krasznahorkay et al. \protect\cite{Sn-skin1,Sn-skin2}. The skin thickness is defined with the deference between neutron and proton rms radii with the equation
$\delta r=\sqrt{<r^2_n>}-\sqrt{<r^2_p>}$. }
\label{FIG1:FIG1}
\end{figure}

\begin{figure}
\includegraphics[height=.26\textheight,width=.42\textwidth]{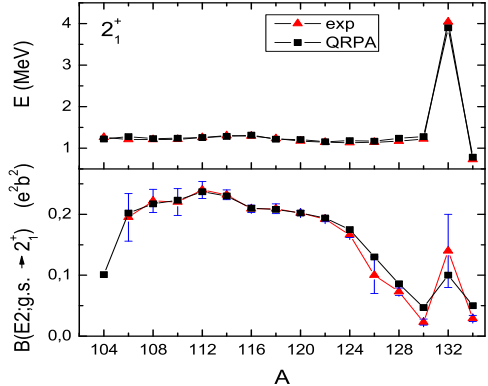}\\
\caption{ QRPA calculations of the energies (upper panel) and transition strengths of the first $2^{+}_{1}$ states in $^{104-134}Sn$ compared to experimental data \cite{Sn106,Sn108,Sn106108,Sn110,Sn112,Sn114,Raman,Radford}.}
\label{FIG2:FIG2}
\end{figure}

\begin{figure}
\includegraphics[height=.18\textheight,width=.41\textwidth]{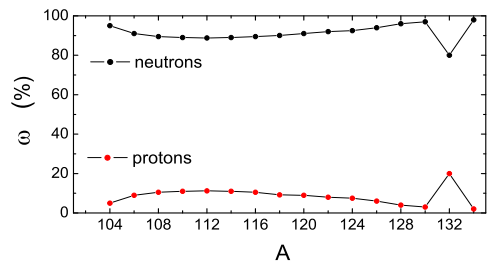}
\caption{ Calculations the neutron and proton contents in the structure of the $2^{+}_{1}$ states in terms of weight factors $\omega_{j_1j_2}({\lambda \mu i}$ where the sum is taken over the neutron or proton components entering in the structure of a phonon.}
\label{FIG3:FIG3}
\end{figure}

In more detail, the residual interactions are given by 
\begin{equation}
H_{res}=H_M^{ph}+H_{SM}^{ph}+H_M^{pp} \quad .
\end{equation}
At present, we do not attempt a fully self-consistent derivation of $H_{res}$ but
follow the empirical approach of the QPM. Hence, we use separable multipole-multipole $H_M^{ph}$ and spin-multipole
$H_{SM}^{ph}$ interactions both of isoscalar and isovector type in the particle-hole (ph)  and multipole pairing $H_M^{pp}$ in the particle-particle (pp) channels, respectively \cite{Sol92}. Further details for the definition of the interactions and the determination of the coupling strength parameters have been discussed extensively e.g. in \cite{Nadia08} and it will not be repeated here. We only mention the interesting observation that in the nuclei considered here the isoscalar quadrupole constant $\kappa^{(2)}_{0}$ is rather independent of the nuclear mass with only slight variations, typically in the order of 1$\%$ between neighboring stable nuclei and less than 5$\%$ outside the valley of stability. This allows safe extrapolations into the hitherto unexplored regions toward $^{100}$Sn and beyond $^{132}$Sn.

\section{Quadrupole excitations in Sn isotopes}
\subsection{The quadrupole response functions}

Before turning to the discussion of the quadrupole response, it is worthwhile to refer again to our previous studies. The HFB calculations \cite{Nadia08} have shown that in the mass region
A$\ge$106, the tin nuclei exhibit a neutron skin while the situation reverses for the lower masses to a tiny proton skin in $^{100-104}Sn$, as also seen in Fig. 1. A neutron skin, being a static property of the nuclear ground state, manifests itself also in the dynamic sector in terms of the low-energy PDR dipole excitations. According to theory, the phenomenon is quite ubiquitous, shown e.g. by our previous studies in the tin isotopes \cite{Nadia04,Nadia08} and the N=50, 82 nuclei \cite{Volz06,Ts-Schw08}. Meanwhile, also a growing evidence from experiments is obtained. In general, the PDR states are of mixed isospin symmetry and their total transition strength is closely related to the size of the neutron skin  \cite{Nadia08}.

Now coming to the present investigations, we first discuss the properties of the traditionally known parts of the quadrupole spectrum. These considerations serve also as tests before focusing on a new sector of the quadrupole spectrum. The QRPA results for energies and B(E2) transition probabilities of the $2^{+}_{1}$ state are displayed and compared to data \cite{Sn106108,Sn110,Sn112} in Fig. 2. The excitation energies are remarkably stable over the mass range, except at the shell closure at A=132. Also the transition strengths vary only mildly in a narrow band for most of the isotopes but decreasing when $^{132}$Sn is approached. In $^{104}$Sn we predict a $2^{+}_{1}$ state with excitation energy $E_x$=1.221~MeV and B(E2;g.s.$\rightarrow$ 2$^{+}$)=0.10 $e^{2}b^{2}$. Thus, we confirm the experimentally observed tendency of decreasing B(E2) values (also found in shell model calculations) in nuclei with extreme proton-to-neutron ratios N/Z toward $^{100}$Sn \cite{Sn106108}.

Here, the region around the double magic nucleus $^{132}$Sn is of particular interest. The restoration of N=82 neutron-shell gap was recently confirmed by high-precision mass measurements \cite{Sn132}. From a systematic study of $2^+_1$ states the authors of \cite{Ter02} found that the product of transition strength and excitation energy is almost constant, i.e. B(E2; g.s.$\rightarrow$2$^{+}_{1}$)$\approx$ 1/E(2$^{+}_{1}$). First of all, this means that the contributions of the $2^+_1$ states to the electromagnetic energy weighted sum rule (EWSR) is to a good approximation a constant. Since the EWSR is a conserved quantity, this puts constraints on the higher lying energy weighted strength which, consequently, is separately conserved as well.

For the explanation of this effect the collectivity of the excited states plays a crucial role. Microscopically, collectivity is directly related to the coherent, i.e. in-phase, excitation of a large number of two-quasiparticle pairs with respect to a transition operator. Favorable candidates are those states which are shifted by interactions into spectral regions away from the poles of the non-interacting two-quasiparticle Green function (i.e. the 4-point function). Namely, more collective states have lower excitation energies and respectively stronger transition rates (because the nucleons contribute coherently). An exception from this rule are nuclei around the shell closures as observed in N=126 (Pb region)\cite{Raman} and N=82 \cite{Ter02}. Our QRPA results for  $^{130-134}$Sn agree with the conclusions drawn in ref. \cite{Ter02,Sev08}.

The abnormal behavior of the B(E2) in nuclei around $^{132}$Sn (see Fig. 2) can be traced back to an increase of the amplitude of the $1g_{9/2}2d_{5/2}$ two-quasiparticle proton component in the first QRPA 2$^{+}$ state in $^{132}$Sn, amounting for about 17\% of the state vector. Compared to the neighboring nuclei $^{130}$Sn and $^{134}$Sn, where this component contributes with probabilities of 1.5\% and 0.9\%, respectively, this is a substantial reordering of strength. Results of the summed proton/neutron amplitudes in terms of weight factors
\begin{equation}
\omega_{j_1j_2}({\lambda \mu i})=\sum|\psi_{j_1j_2}^{\lambda\mu
i}|^2-|\varphi_{j_1j_2}^{\lambda\mu i}|^2*100 [\%],
\label{norm}
\end{equation}
contributing to the state vector of the 2$^{+}_{1}$ state in Sn isotopes are presented in Fig. 3.

\begin{figure*}
\includegraphics[height=.4\textheight]{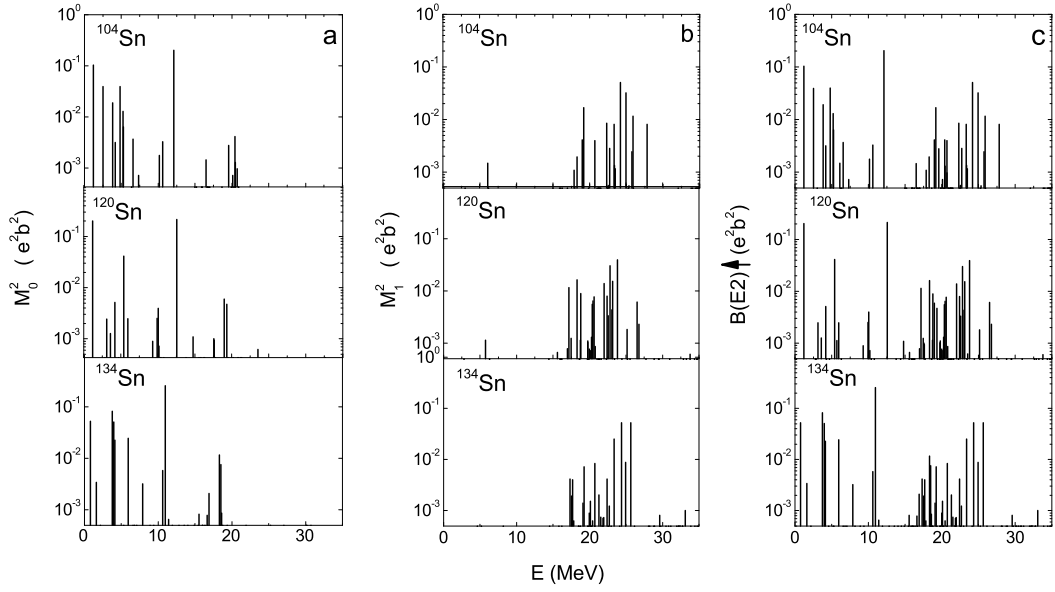}
\caption{Calculations of isoscalar (a) and isovector (b) electric quadrupole strengths in comparison with the total QRPA B(E2) transition strengths (c) up to 35 MeV in Sn isotopes.}
\label{FIG4:Fig4}
\end{figure*}
\begin{figure*}
\includegraphics[height=.4\textheight]{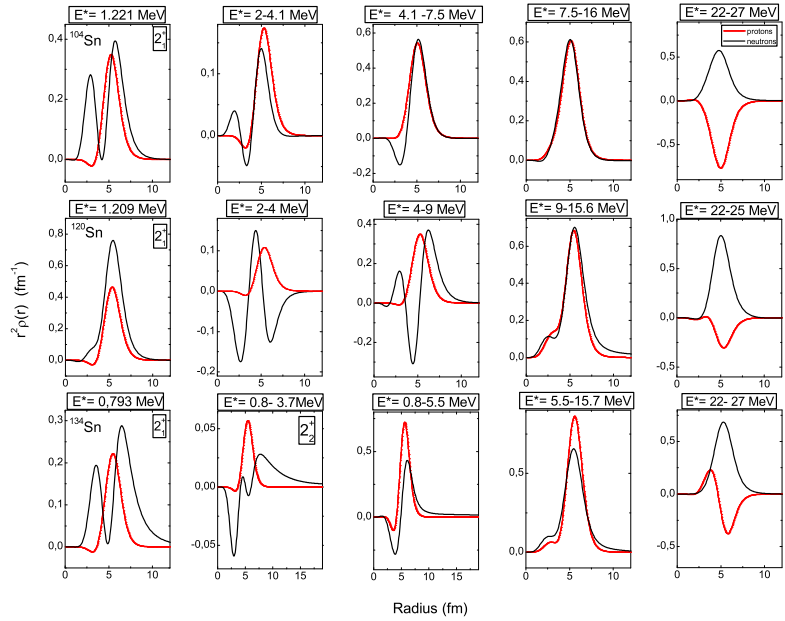}
\caption{QRPA proton (dash line) and neutron (solid line) quadrupole transition densities summed over 2$^+$ excited states in a given energy region (indicated over every plot) in $^{104,120,134}$Sn isotopes.}
\label{FIG5:FIG5}
\end{figure*}

After having fixed accurately the model parameters and since both energies and B(E2)-values of the lowest quadrupole states are well reproduced, we expect that the calculation will lead to reliable results also for the quadrupole states at intermediate excitation energies. Our particular interest is to investigate whether there is a correlation between the neutron (proton) skin and quadrupole states in the energy region below and around the particle separation threshold, similar to the dipole case. For completeness and as further test cases we also include the higher-lying collective states up to the isovector giant quadrupole resonances (IVGQR) \cite{Hara} and above until $E_x=35$~MeV.
Similarly to the QRPA results for the PDR modes, in Sn isotopes (N= 66-84) we observe a sequence of quadrupole states with common spectroscopic features, located just above the collective 2$^{+}_1$ and below the ISGQR.  The lowest-lying of them, located typically in the energy region $E_x\approx$ 2-4 MeV (with exception the lighter tins and $^{132}$Sn), are non-collective, with predominantly neutron structure (the proton contribution is less than a few percents). With increase of the excitation energy above $E_x$ $\approx$ 4 MeV toward the particle emission threshold, the collectivity of the states increases, though they have still more pronounced neutron components for A$\geq$106. 
    
We suggest that the low-energy quadrupole excitations with predominantly neutron structure in tin isotopes could belong to a new quadrupole mode, namely, a neutron PQR which manifests itself as a neutron surface vibrations. A supporting argument in this direction could be a possible connection between the B(E2) transition rates of the PQR states and the neutron number of the nucleus related to the thickness of the neutron skin \cite{Nadia08}. 

From systematic studies of B(E2) transitions of quadrupole states with excitation energies $E_x\approx$ 2-4 MeV in the tin isotopes we find a correlation between the summed over non-collective neutron modes B(E2) transitions and the number of the neutrons of the nucleus related to the neutron skin thickness. Theoretical calculations of B(E2) spectral transition strength distributions in $^{104,120,134}$Sn isotopes are presented in Fig. \ref{FIG4:Fig4}. A sizable effect of increased B(E2) at $E_x\approx$ 2-4 MeV is observed for the heaviest tin isotopes - $^{130}$Sn and $^{134}$Sn.
For $^{132}$Sn the neutron excitations in this energy range are suppressed due to the N=82 shell closure.
In the lighter tin isotopes the reduced neutron number decreases the collectivity of the 2$^{+}_{1}$ state which becomes non-collective in $^{104}$Sn. At the same time the proton contribution to the B(E2) strength located in the  energy range  2-4 MeV increases toward $^{104}$Sn and brings to more intensive proton quadrupole excitations in $^{104}$Sn there (see Fig. \ref{FIG4:Fig4}c). Consequently, the $^{104}$Sn nucleus appears to be an opposing point where the neutron skin chances to proton and the neutron PQR to proton PQR, respectively. 
The investigations of the nature of the PQR excitations in $^{104-134}$Sn nuclei reveal that these states are predominantly of isoscalar character in the heavier tin isotopes. The contribution of isovector strength in $^{130}$Sn and $^{134}$Sn in the PQR region is negligibly small but it increases toward lighter tin isotopes due to the larger proton content of the state vectors. The calculations of the distributions of the isoscalar M$_0$ and isovector M$_1$ quadrupole strengths in $^{104-134}$Sn up 35 MeV are presented in Fig. \ref{FIG4:Fig4}a,b, respectively.
The isoscalar (isovector) electric matrix elements M$_{0(1)}(\lambda\mu)$ for $\lambda = 2$ are defined with the equation:
\begin{equation}
M_{0(1)}(2^+)\approx \left\langle 2^+\left\|\sum^{p}_{k}r^{2}_{k}Y_{2 \mu}(\Omega_{k})^\pm 
\sum^{n}_{k}r^{2}_{k}Y_{2 \mu}(\Omega_{k})
\right\|g.s.\right\rangle
\label{iso}
\end{equation}    

The sign '+' in this equation corresponds to the isoscalar case for which the entering 2$^+$ states have the same signs of the proton and neutron summed single-particle matrix elements. The sign '-' is related to the isovector case where the proton and neutron summed single-particle matrix elements are with opposite signs, respectively.

\subsection{Quadrupole Transition Densities} 

The characteristics of the PQR states are seen most clearly by inspection of transition densities, defining the non-diagonal parts of the one-body density matrix \cite{Nadia08}. In this way, we expect a deeper insight into the spatial pattern and location of the transition strength, similar to the analysis for dipole excitations \cite{Nadia08}. In Fig. \ref{FIG5:FIG5} results for the proton and neutron transition densities of the isoscalar 2$^{+}_{1}$ state in several tin isotopes are displayed. From the QRPA state vectors one finds that the 2$^{+}_{1}$ state changes from collective in the middle of the shell (between N=50 and N=82) to non-collective toward the shell closure.

In $^{106-134}$Sn most of the quadrupole states below the particle emission threshold are of isoscalar character. However, some mixed symmetry configurations, indicated by the opposite signs of the proton and neutron amplitudes, are also observed in this region, as seen in the third $^{120}$Sn panel, Fig.\ref{FIG5:FIG5}. At lower energies, $E_x$=2-4 MeV and 
$E_x$=0.8-3.7 MeV in $^{120}$Sn and $^{134}$Sn, respectively 
the main part of the oscillations is coming from less strongly bound neutrons, forming a skin-like surface layer \cite{Nadia08}, located predominantly at the nuclear periphery, where the nuclear radii are extended above an average value of R$\sim$5~fm up to R$\approx$10 fm in $^{120}$Sn and up to R$\approx$20 in $^{134}$Sn. At the same time we find a very small proton contribution in this space region. 

For the double-magic $^{132}$Sn the first QRPA 2$^{+}$ ($E_x$=3.9 MeV) and the second QRPA 2$^{+}$ ($E_x$= 5.3 MeV) states exhibit properties similar to the neutron states discussed in the other tin isotopes in the same energy range. These states incorporate strength related to oscillations of neutrons at the surface region. In addition in the second 2$^{+}$ state in $^{132}$Sn in the nuclear interior (R $\leq$ 5 fm) the protons and neutrons vibrate in-phase where proton an neutron parts are of comparable amplitude.

With change in the structure of the quadrupole states at $E_x$=2- 5 MeV in respect of increase in the proton contribution in the states vectors toward $^{104}$Sn the proton and neutron transition densities undergo a change in their behavior as it is demonstrated in Fig. \ref{FIG5:FIG5}. The proton oscillations at the nuclear surface at $E_x$= 2- 4.3 MeV in $^{104}$Sn prevail over the neutron oscillations.
The observations of the change of the behavior of the transition densities
in $^{104}$Sn could be a sign for transformation of the character of the PQR from neutron to proton (similarly to the PDR case \cite{Nadia08,Paar05,Bar08}).

With the increase of the excitation energy $E_x$ toward the particle separation threshold, the quadrupole states become more collective with a larger admixture of isovector components to the state vectors.
Nevertheless, some isospin effects are still present up to the particle emission threshold (even strongly hindered).
In this case, in $^{106-130}$Sn isotopes we observe more isovector type of oscillations inside the nucleus while at the surface neutrons dominate. 
The nature of these higher-lying states to our opinion could not be related directly to skin effects as they collect a lot of strength (some of the them have transition probability B(E2)$\uparrow$ of order of 10 W.u., which is of the same size as for the collective 2$^{+}_{1}$ and the IS(V)GQRs) and the contribution of nucleons from the nuclear interior is significant. At the same time the role of the isospin effects, in particular the neutron skin for energies close to the threshold has to be further investigated.
The last two plots (down) in every tin nucleus presented in Fig. \ref{FIG5:FIG5} show the proton and neutron transition densities of the ISGQR and IVGQR, respectively.

Overall, the transition densities show an interesting dependence on the excitation energy: in all nuclei the isoscalar character of the lowest $2^+$ state and the GQR is evident from the in-phase behavior of proton and neutron components, while proton and neutron transition densities carry opposite phases in the isovector GQR region. Together with the almost identical radial shapes these results follow almost perfectly well the generally accepted rules for collective quadrupole states. They are well interpreted as surface and volume oscillations of a liquid drop \cite{Rowe}. However, the transition densities related to the additional quadrupole states which may be considered as PQR excitations are not following any of the known rules but are showing quite unusual properties. As seen from Fig. \ref{FIG5:FIG5} the proton components continue to behave like vibrational variations of the nuclear density radius while the neutron transition densities develop a rather different nodal pattern. The nodal structures correspond to processes where a (tiny) portion of nuclear matter is shuffled around the nuclear radius, leaving the latter almost unaffected, as indicated by the radial node occurring at or close to the nuclear half density radius, $R\sim 5$~fm. In a liquid drop picture, these features can be interpreted as vibrational excitations of the nuclear surface given by a superposition of variations in position and diffuseness of the nuclear surface. The latter mode was already anticipated by Bohr and Mottelsen \cite{BM2}, although hitherto experimentally never verified, at least not in stable nuclei. Our results give evidence that such modes may indeed exist in charge-asymmetric nuclei.

This interpretation is supported by a result of Pethick and Ravenhall who derived in ref. \cite{Pet96} a relation connecting the thickness of the nuclear skin to the surface tension. According to their result, the surface tension decreases with increasing neutron excess. This is a statement concerning the static properties of the nuclear ground state density which is at least qualitatively fully supported by our microscopic HFB ground state. densities. Now, as mentioned, the shapes of the PQR neutron transition densities are compatible with a reordering of neutron skin matter around the nuclear surface, hence inducing a local variation of the neutron excess in the surface region, resulting in oscillatory alterations of the surface tension.

\section{Conclusions}

Quadrupole states were investigated in $^{104-134}$Sn nuclei up to excitation energies of 35 MeV. From QRPA calculations of 2$^{+}_{1}$ states, the QPM model parameters are reliable determined. For the double-magic $^{132}$Sn an increase of the $B(E2; g.s. \rightarrow 2^{+}_{1})$ compared to the neighbor $^{130}$Sn and $^{134}$Sn is observed and explained. For $^{104}$Sn, the first 2$^{+}$ state is predicted and its spectroscopic properties are determined.
From the analysis of isoscalar and isovector electric quadrupole strength distributions the nature of the quadrupole states is investigated and 
a separation between collective isoscalar, isovector and low-energy mixed symmetry states is achieved. The spectral distribution of the low-energy mixed symmetry states in $^{104-134}$Sn are clustering in a confined energy region such that they may be considered forming a PQR.
An increase of the lowest-energy non-collective strength toward $^{134}$Sn
due to mainly neutron excitations is found. 
In addition, the correlation of the PQR transition strength with the neutron (proton) skin thickness manifests itself via a transition from a neutron PQR to a proton PQR in $^{104}$Sn, the mass region where the neutron skin reverses into a proton skin.
By means of quadrupole transition densities the nature of the PQR is clarified.
In general, the PQR resembles the properties of the PDR and could be related to skin oscillations of one type of nucleons. 
Even though, with the increase of the collectivity of the quadrupole excitations the reliable distinction of the pure skin neutron (proton) quadrupole oscillations seems to be a difficult task, especially for nuclei with small or moderate neutron excess. In this respect, nuclei with larger isospin asymmetry like $^{130}$Sn and $^{134}$Sn could be a good candidates for experimentally explorations. 

Supported by DFG project Le 439/1-4 and BMBF.

\end{document}